\documentclass{article}

\usepackage{amsmath,amsfonts,setspace,cancel,url,hyperref,verbatim,color,todonotes,epsfig}
\usepackage{cite}
\usepackage{comment}
\usepackage[a4paper]{geometry}
\usepackage{graphicx}

\renewcommand{\phi}{\varphi}
\renewcommand{\theta}{\vartheta}

\newcommand{\be}{\begin{equation}}
\newcommand{\ee}{\end{equation}}
\newcommand{\ba}{\begin{array}}
\newcommand{\ea}{\end{array}}
\newcommand{\bea}{\begin{eqnarray}}
\newcommand{\eea}{\end{eqnarray}}
\newcommand{\gL}{\mathcal{L}}
\newcommand{\tpartial}{\tilde{\partial}}
\newcommand{\hpartial}{\hat{\partial}}
\newcommand{\SL}[1]{\mathrm{SL}( #1 )}
\newcommand{\SLf}{\SL{5}}
\newcommand{\SOf}{\mathrm{SO}(4)}
\newcommand{\SUt}{\mathrm{SU(2)}}
\newcommand{\Oct}{\mathbb{O}}
\newcommand{\x}{x^1}
\newcommand{\y}{x^2}
\newcommand{\z}{x^3}
\newcommand{\w}{x^4}
\newcommand{\dx}{d\x}
\newcommand{\dy}{d\y}
\newcommand{\dz}{d\z}
\newcommand{\dw}{d\w}
\newcommand{\pdx}{\left(\dx\right)}
\newcommand{\pdy}{\left(\dy\right)}
\newcommand{\pdz}{\left(\dz\right)}
\newcommand{\pdw}{\left(\dw\right)}

\numberwithin{equation}{section}
\begin{document}

\begin{titlepage}
\vfill

\begin{flushright}
IGC-16/7-1 \\
MPP-2016-163 \\
LMU-ASC 33/16
\end{flushright}

\vfill

\begin{center}
   \baselineskip=16pt
   	{\Large \bf Non-associativity in non-geometric string and M-theory backgrounds, the algebra of octonions, and missing momentum modes}
   	\vskip 2cm
	{\bf Murat G\"unaydin$^{a}$\footnote{\tt mgunaydin@psu.edu}, Dieter L\"ust$^{b,c}$\footnote{\tt dieter.luest@lmu.de}, Emanuel Malek$^{b}$\footnote{\tt e.malek@lmu.de}}
	\vskip .6cm
    {\small \it $^a$ Institute for Gravitation and the Cosmos, Physics Department, \\ Pennsylvania State University, University Park, PA 16802, USA \\ \ \\
    $^b$ Arnold Sommerfeld Center for Theoretical Physics, Department f\"ur Physik, \\ Ludwig-Maximilians-Universit\"at M\"unchen, Theresienstra{\ss}e 37, 80333 M\"unchen, Germany \\ \ \\
    $^c$ Max-Planck-Institut f\"ur Physik, Werner-Heisenberg-Institut, \\ F\"ohringer Ring 6, 80805 M\"unchen, Germany \\ \ \\}
	\vskip 2cm
\end{center}

\begin{abstract}
We propose a non-associative phase space algebra for M-theory backgrounds with locally non-geometric fluxes based on the non-associative algebra of octonions. Our proposal is based on the observation that the non-associative algebra of the non-geometric $R$-flux background in string theory can be obtained by a proper contraction of the simple Malcev algebra generated by  imaginary octonions. Furthermore, by studying a toy model of a four-dimensional locally non-geometric M-theory background which is dual to a twisted torus, we show that the non-geometric background is ``missing'' a momentum mode. The resulting seven-dimensional phase space can thus be naturally identified with the imaginary octonions. This allows us to interpret the full uncontracted algebra of imaginary octonions as the uplift of the string theory $R$-flux algebra to M-theory, with the contraction parameter playing the role of the string coupling constant $g_s$.
\end{abstract}

\vfill

\setcounter{footnote}{0}
\end{titlepage}

\tableofcontents

\newpage

\section{Introduction}

Flux backgrounds received considerable attention in recent years, while developing superstring theory and its viable phenomenological applications to model building for elementary particle physics (for reviews see \cite{Grana:2005jc,Blumenhagen:2006ci}). T-duality transformations in the presence of fluxes, such as Neveu-Schwarz fluxes associated to a closed 3-form $H$, were used to exhibit that not only the geometry but also the topology of space in which strings propagate are not perceived as in ordinary point-particle theories, since T-duality relates backgrounds with different topologies.

As an important example, T-duality of string flux backgrounds that are $\mathrm{U}(1)$-fibrations over a base manifold $M$ provides non-geometric flux backgrounds, which in turn lead to new phenomena like  non-commutativity and non-associativity of coordinates in the context of closed string theory       \cite{Lust:2010iy,Blumenhagen:2010hj,Blumenhagen:2011ph}. In particular, for $n=2$ (hereafter called the non-geometric $Q$-flux background), the resulting closed string background is only locally geometric, but not globally, since the transition functions between two coordinate patches are prescribed in terms of T-duality transformations and not in terms of diffeomorphisms. Then, because of non-trivial monodromies characterizing the $T^2$ fibration over $S^1$ in the $Q$-flux model, the coordinates become non-commutative and the appropriate mathematical structure is that of a non-commutative 2-torus fibred over $S^1$. For $n = 3$ (called the non-geometric $R$-flux background) the situation becomes even more interesting as non-geometric closed string $R$-flux backgrounds exhibit a non-associative structure. Here the resulting closed string $R$-flux background not only fails to be globally geometric, but also locally.  Similar results based on somewhat less explicit methods appeared in the literature before \cite{Bouwknegt:2003vb,Bouwknegt:2003zg,Bouwknegt:2004ap,Bouwknegt:2004tr,Mathai:2005fd}, while studying the action of topological T-duality on torus fibrations with fluxes.

The emergence of closed string non-commutativity and non-associativity was shown in explicit string and CFT models \cite{Lust:2010iy,Blumenhagen:2011ph,Condeescu:2012sp,Andriot:2012vb,Blair:2014kla,Bakas:2015gia}, where the non-geometric structures are due to left-right asymmetric world-sheet constructions, which are very similar to asymmetric orbifold compactifications. Furthermore the non-associative deformation of geometry was also discussed in the context of double field theory \cite{Blumenhagen:2013zpa}, where non-associativity arises after some violation of the strong constraint by the $R$-flux background geometry. Finally the non-associative $R$-flux algebra was also derived from a membrane sigma-model \cite{Mylonas:2012pg,Mylonas:2014aga}. Here we are going to focus on the prototypical non-geometric models, namely the so-called parabolic non-geometric string $R$-flux models on a three-torus with constant fluxes, where we have the following non-trivial commutation relations among the closed string coordinates and momenta
\begin{equation}
\left[ x^i, x^j \right] = i \frac{l_s^3}{\hbar} R^{ijk} p_k \,,
\label{commclosed}
\end{equation}
whereas the remaining commutation relations retain their standard form,
\begin{equation}
\left[x^i, p^j\right]=i \hbar \delta^{ij}\,, \quad \left[p^i, p^j\right]=0 \,.
\label{commclo23}
\end{equation}
It follows that the Jacobiator among the closed string coordinates is non-vanishing:
\begin{equation}
\left[ x^i, x^j, x^k\right] \equiv \frac13 \left[\left[x^1, x^2\right], x^3\right]+{\rm cycl.~perm.} = l_s^3 R^{ijk}\, ,
\label{asso}
\end{equation}
which demonstrates that the underlying algebra is not only non-commutative but also intrinsically non-associative.

We should note that non-commutativity and non-associativity have been known to arise in the description of electrically charged particles moving in the field of magnetic charge. As was shown a long time ago the commutator of velocities of an electron moving in the field of a point-like magnetic monopole do not satisfy the Jacobi identities at the point of the monopole \cite{Lipkin:1969ck}. The mathematical description of charged particles moving in the field of magnetic charge involves nontrivial three-cocycles which were studied in \cite{Grossman:1984fs,Jackiw:1984rd,Jackiw:1985hq,Wu:1984wr,Mylonas:2012pg,Bakas:2013jwa}. For a charged particle moving in the field of constant magnetic charge distribution the velocities and coordinates form a non-Lie Malcev algebra as was shown in \cite{Gunaydin:1985ur}, and which the authors referred to as a ``magnetic  algebra''. This implies that the corresponding quantum mechanical description of the magnetic algebra can not be achieved in terms of linear operators acting on a Hilbert space which are necessarily associative. More recently it was pointed out that the magnetic algebra of \cite{Gunaydin:1985ur} is isomorphic to the R-flux algebra for constant R-flux with the roles of coordinates and momenta interchanged \cite{Gunaydin:2013nqa,Bakas:2013jwa}.

On the mathematical side, the algebra of octonions is a prime example of a non-associative algebra, which however has so far not been related to the non-associative $R$-flux algebra displayed above or the magnetic algebra of \cite{Gunaydin:1985ur}. The algebra of octonions, which we briefly review in appendix B,  is closely related to already known associative structures in M-theory, in particular to the compactification on a seven sphere with torsion whose isometry groups contain the automorphism group $G_2$ of octonions \cite{Gunaydin:1983mi,deWit:1983gs}, to compactifications on $G_2$-holonomy manifolds and to some of the non-associative three-algebras, which were found in the context of multiple membranes in M-theory as well as in ${\cal N}=6$ Chern-Simon gauge theories in three dimensions.

Furthermore, the continuous U-duality groups of maximal supergravity\cite{Cremmer:1979up} and exceptional supergravity \cite{Gunaydin:1983rk} in $d=5,4$ and $3 $ dimensions are the exceptional groups of the E-series that can be realised geometrically  via the exceptional Jordan algebra and its associated Freudenthal triple system defined over the split and real octonions, respectively\footnote{ The exceptional Jordan algebra of $3\times 3$ Hermitian octonionic matrices is the unique Jordan algebra that has no realization in terms of associative matrices with the symmetric Jordan product taken as the anti-commutator. We refer to \cite{Gunaydin:2000xr,Gunaydin:2005zz} and the review  \cite{Gunaydin:2009pk} and references therein for geometric realizations of exceptional groups over  Jordan algebras and Freudenthal triple systems. }.

The purpose of our paper is two-fold: first we will demonstrate that the non-associative $R$-flux algebra in eqs. \eqref{commclosed}-\eqref{asso} is indeed very closely related to the algebra of octonions $\Oct$. More specifically we will show that a suitable contraction of the non-commutative and non-associative algebra generated by the seven imaginary units $e_A$ of octonions precisely reproduces the algebra \eqref{commclosed}-\eqref{asso}.

Given that the non-associative string theory $R$-flux algebra is obtained by a contraction of the Malcev algebra of imaginary octonions, we then ask what role the full uncontracted algebra plays. Our second goal is to propose that the uplift of the string $R$-flux algebra to M-theory is described by the simple Malcev algebra generated by the uncontracted division algebra of octonions. We will show, in particular, that the resulting uplift is compatible with the uplift of the non-geometric fluxes to M-theory within the context of exceptional field theory \cite{Blair:2014zba}.

We substantiate this proposal further by studying a four-dimensional toy model for an $R$-flux background which is dual to a twisted torus. We show that the locally non-geometric background is ``missing'' a momentum mode and as a result the phase space is seven-, not eight-dimensional as would be naively expected. This allows us to identify the phase space variables, four coordinates and three momenta, with the seven imaginary units of octonions. 

In this way we provide evidence that M-theory with non-geometric fluxes also exhibits non-associativity among its coordinates which correctly reduces down to the string theory non-associative algebra when performing a suitable contraction. The parameter which we introduce for the contraction of the full octonionic algebra takes within the M-theory uplift the role of the string coupling constant $g_s$, in other words the radius of the circle in the 11th direction of M-theory.

The present paper is organised as follows. Starting from T-duality and non-geometric fluxes on the three-torus $T^3$, we will review in the next chapter how the non-geometric fluxes get uplifted to M-theory on four-dimensional spaces. In particular this uplift is described using the $\mathrm{SL}(5)$ exceptional field theory, which extends the $\mathrm{Spin}(3,3) \simeq \mathrm{SL}(4)$ double field theory of string theory on three-dimensional spaces. We then discuss in section \ref{s:Contraction} how the non-associative string $R$-flux algebra can be obtained by a contraction of the algebra generated by imaginary octonions under commutation. In section \ref{s:MNonAssoc} we conjecture that the full uncontracted algebra of imaginary octonions provides the uplift of the string $R$-flux algebra to M-theory. In particular, we revisit the four-dimensional locally non-geometric M-theory toy-model to show that a momentum mode is missing and hence that the phase space is seven-dimensional. This allows us to identify the seven imaginary octonions with the phase space variables and provides evidence for our conjecture. Finally, we conclude in section \ref{s:Discussion}.

\section{The $R$-flux background and its uplift to M-theory and exceptional field theory} \label{s:ToyModels}
In order to give the reader a feel for what a locally non-geometric background is we start by reviewing a toy model for such backgrounds in string theory \cite{Kachru:2002sk,Hull:2004in,Shelton:2005cf,Dabholkar:2005ve} and its uplift to M-theory \cite{Blair:2014zba}. This toy model will also guide us when we look to generalise the non-associative algebra for locally non-geometric M-theory backgrounds in section \ref{s:MNonAssoc}.

\subsection{The $T^3$ duality chain in string theory} \label{s:StringToyModel}
Let us start by recalling the standard chain of dualities for string theory on $T^3$ \cite{Kachru:2002sk,Hull:2004in,Shelton:2005cf,Dabholkar:2005ve}:
\begin{equation}\label{eq:TdualityChain}
H_{ijk} \stackrel{T_{i}}{\longrightarrow} f^{i}_{jk}
\stackrel{T_{j}}{\longrightarrow} Q_{k}^{ij}
\stackrel{T_{k}}{\longrightarrow} R^{ijk}\, ,\quad (i,j,k=1,\dots , 3)\,,
\end{equation}
where $T_i$ denotes a T-duality along the $x^i$ direction. Note that only the first background describes a torus, with the other backgrounds being more general $\mathrm{U}(1)$-fibrations, which in the latter two cases can be defined over the ``doubled space'' of double field theory.

\paragraph{$T^3$ with $H$-flux:}
The $T^3$ duality chain starts with a 3-torus with $H$-flux
\begin{equation}
 ds^2 = \pdx^2 + \pdy^2 + \pdz^2 \,, \qquad B_{12} = N \z \,, \label{eq:HFlux}
\end{equation}
where the Kalb-Ramond two-form $B_{12}$ is not globally well-defined and hence there is a non-zero $H$-flux through the $T^3$ given by
\begin{equation}
 H_{123} = N \,.
\end{equation}

\paragraph{Twisted torus:}
We now perform a duality along the $\x$ direction and obtain a twisted torus with no $H$-flux.
\begin{equation}
 ds^2 = \left( \dx - N\z\dy \right)^2 + \pdy^2 + \pdz^2 \,, \qquad B_{2} = 0 \,. \label{eq:Twist3Torus}
\end{equation}
We denote the twisted torus as $\tilde{T}^3$, where the coordinate $x^1$ of the twisted torus corresponds to the dual coordinates $\tilde x_1$ of the $H$-flux background. The coordinates of the twisted torus are to be identified as
\begin{equation}
 \left( \x,\y,\z \right) \sim \left( \x + 1,\y,\z \right) \sim \left( \x,\y + 1,\z \right) \sim \left( \x + N \y,\y,\z + 1 \right) \,.
\end{equation}
One can understand this as follows: while in the $H$-flux background, $B$ gets patched by a gauge transformation, after a T-duality this patching gets shifted into the diffeomorphism group of the torus. As a result the twisted torus $\tilde{T}^3$ can be viewed as a $\mathrm{U}(1)$-bundle over $T^2$ with non-vanishing first Chern class. Alternatively, since $\tilde{T}^3$ is parallelisable we can describe the ``twist'' using the spin connection. Denoting the three globally well-defined one-forms as
\begin{equation}
 \eta^1 = \dx - N\z\dy \,, \qquad \eta^2 = \dy \,, \qquad \eta^3 = \dz \,, \label{eq:TwistedTorus1Forms}
\end{equation}
we find
\begin{equation}
 d\eta^1 = N \eta^2 \wedge \eta^3 \,, \qquad d\eta^2 = d\eta^3 = 0 \,.
\end{equation}
As a result the spin connection, defined as
\begin{equation}
 d\eta^i = \omega^i_{jk} \eta^j \wedge \eta^k \,,
\end{equation}
becomes
\begin{equation}
 \omega^{1}_{23} = N \,,
\end{equation}
and we see that the $H$-flux has turned into the ``geometric flux'' which we here identify with the spin-connection $f^i_{jk} = \omega^i_{jk}$.

\paragraph{$Q$-flux background:}
Performing another T-duality in the $\y$-direction we get a background that is not globally well-defined. The metric and two-form would be given as
\begin{equation}
 ds^2 = \frac{\pdx^2 + \pdy^2}{1+N^2\left(\z\right)^2} + \pdz^2 \,, \qquad B_{23} = \frac{N\z}{1+N^2\left(\z\right)^2} \,. \label{eq:QFlux}
\end{equation}
Now the coordinate $x^2$ of the $Q$-flux background corresponds to the dual coordinates $\tilde x_2$ of the twisted torus.
This is clearly not well-defined in conventional geometry but instead the background is patched with an element of the $\mathrm{SO}(3,3)$ duality group as $\z \rightarrow \z+1$. Such a background is called a T-fold and is expected to belong to a class of permissible string backgrounds.

The metric and Kalb-Ramond two-form are actually not the correct variables to describe the Q-flux background \eqref{eq:QFlux} since they are not globally well-defined. Instead, the background can be expressed in terms of a bivector $\beta^{ij}$ which is well-defined, and is related by a field redefinition \cite{Duff:1989tf,Grana:2008yw,Andriot:2011uh,Andriot:2012wx,Andriot:2012an,Blumenhagen:2013aia,Andriot:2013xca} (or equivalently in the context of generalised geometry and double field theory an $\mathrm{O}(3) \times \mathrm{O}(3)$ rotation which relates different parameterisations of the generalised vielbein)
\begin{equation}
 \begin{split}
  \beta^{ij} &= \frac12 \left( \left( g - B \right)^{-1} - \left( g+B \right)^{-1} \right) \,, \\
  \hat{g} &= \frac12 \left( \left(g - B\right)^{-1} + \left(g + B\right)^{-1} \right)^{-1} \,. \label{eq:betaFieldRedefinition}
 \end{split}
\end{equation}
The metric and bivector are then given by
\begin{equation}
 \hat{ds}^2 = \pdx^2 + \pdy^2 + \pdz^2 \,, \qquad \beta^{12} = N\z \,, \label{eq:QFluxBeta}
\end{equation}
with $\hat{ds}^2$ the line element of $\hat{g}$. The field redefinition \eqref{eq:betaFieldRedefinition} is of course not globally well-defined and there is a clearer way to see that the ``non-geometric frame'' with the bivector $\beta$ is preferred, as we briefly discuss in \ref{s:DFTQFluxBackgrounds}. Finally, we wish to note that this field redefinition does not somehow ``cure'' the non-geometry and that as a result the resulting theory cannot simply be thought of as a gravitational theory with some different matter.

This background is classified by its ``$Q$-flux'' which is a spacetime tensor \cite{Andriot:2012wx,Andriot:2013xca} and is in this case given by
\begin{equation}
 Q_i^{jk} = \partial_i \beta^{jk} \,. \label{eq:QFluxDef}
\end{equation}
For the background \eqref{eq:QFluxBeta} its only non-zero components are
\begin{equation}
 Q_3^{12} = N \,.
\end{equation}
Again, we can see that the duality along the $\y$-direction has pushed the patching from the diffeomorphism group into an element of $\mathrm{SO}(3,3)$ which is not an element of the geometric subgroup of diffeomorphisms and gauge transformations. We discuss a clear way to see this from a generalised geometry \cite{Grana:2008yw} and double field theory point of view in appendix \ref{s:DFTQFluxBackgrounds}.

Finally, it is important to emphasise the importance of performing a duality along the $x^2$-direction, where $x^2$ is not a globally well-defined coordinate. Thus $\partial_{2}$ is not a globally well-defined vector field. If we had chosen a globally well-defined vector-field the duality would not have resulted in a non-geometric background.

\paragraph{$R$-flux background:}
Finally, we can perform a duality along the remaining direction, $\z$. This is no longer an isometry and so the usual Buscher procedure fails. However, double field theory gives a framework in which this ``generalised T-duality'' \cite{Dabholkar:2005ve} can be made sense of.\footnote{Recent progress has been made in understanding such generalised T-dualities from non-linear sigma models \cite{Chatzistavrakidis:2015lga,Chatzistavrakidis:2016jci,Chatzistavrakidis:2016vsy}.} It boils down to applying the Buscher rules as if we had dualised along an isometry and also exchanging the coordinate $\z$ with its ``dual coordinate'' $\tilde{x}_3$. In the new coordinate system we would thus have that ${\z} \longrightarrow \tilde{x}_3$ is a dual coordinate on which the background will now depend.

After duality the metric and $\beta$-field are given by
\begin{equation}
 \hat{ds}^2 = \left(dx^2\right)^2 + \left(dx^2\right)^2 + \left(dx^3\right)^2 \,, \qquad \beta^{12} = N \tilde{x}_3 \,. \label{eq:RFluxT3}
\end{equation}
The $R$-flux is defined by \cite{Andriot:2011uh,Andriot:2012wx}
\begin{equation}
 R^{ijk} = 3 \hat{\partial}^{[i} \beta^{jk]} \,,
\end{equation}
with $\hat{\partial}^{i} = \tilde{\partial}^{i} + \beta^{ij} \partial_j$ and is here given by
\begin{equation}
 R^{123} = N \,.
\end{equation}

The backgrounds given here do not define CFTs and hence serve only as a toy-model. However, we expect these effects to be realised in CFTs as well because the non-geometry arises by dualising along a vector field, here $\partial_2$ which is not globally-well defined. This is what causes the dual theory to be ``non-geometric''. In the language of double field theory, the corresponding section which is parameterised by the coordinates $\x$, $\y$ and $\tilde{x}_3$, is not globally well-defined and this generates the non-geometry. Similarly, the $R$-flux background arises because of the dualisation along a direction which is not an isometry. Starting from the twisted torus, the $R$-flux background is generated by dualising along one direction which is not globally well-defined and one which is not an isometry.

In fact, the so-called exotic branes of de Boer and Shigemori \cite{deBoer:2010ud,deBoer:2012ma} and the ``$Q$-branes'' and ``$R$-branes'' \cite{Hassler:2013wsa,Bakhmatov:2016kfn} realise the above duality chain at the level of supergravity solutions, see for example \cite{Blair:2014zba}. Furthermore, it was shown in \cite{Condeescu:2013yma} that in specific left-right asymmetric orbifold constructions the non-geometric $R$-flux emerges in the context of the gauged supergravity algebra after dimensional reduction.

\subsection{Uplift of the string theory duality chain} \label{s:MToy}
Now we wish to describe how these fluxes and the associated T-duality transformations are embedded in M-theory or, respectively, exceptional field theory. The M-theory uplift of the above string compactifications is described by the $\SLf$ exceptional field theory, which governs compactifications to seven dimensions. The non-geometric fluxes of the $\SLf$ exceptional field theory were described in \cite{Blair:2014zba} and we follow the notation of that paper.

If we were considering only IIA backgrounds (the argument can of course be repeated for IIB), we would require two T-dualities so that the above duality chain splits into
\begin{equation}\label{eq:TdualityChainSplit}
H_{ijk} \stackrel{T_{ij}}{\longrightarrow} Q_k^{ij} \,, \qquad f^i_{jk}
\stackrel{T_{jk}}{\longrightarrow} R^{ijk}\,.
\end{equation}
When we uplift to M-theory, two T-dualities become three U-dualities with the third duality along the M-theory circle, ensuring the right dilaton shift. Thus when considering 11-dimensional backgrounds, we need to perform three U-dualities. Then in order to obtain the 11-dimensional analogue of the $R$-flux background we need to act with three U-dualities on a background with geometric flux. Let us thus consider the internal space of a twisted torus times a circle, $\tilde{T}^3 \times S^1$, so that
\begin{equation}
 ds_4^2 = \pdy^2 + \pdz^2 + \pdw^2 + \left(\dx - N\z\dy \right)^2 \,, \qquad C_3 = 0 \,, \label{eq:TwistedTorusMetric}
\end{equation}
with the identifications
\begin{equation}
 \begin{split}
  \left( \x,\y,\z,\w \right) &\sim \left( \x + 1,\y,\z,\w \right) \sim \left( \x,\y + 1,\z,\w \right) \sim \left( \x + N \y,\y,\z + 1,\w \right) \\
  & \sim \left( \x,\y,\z,\w + 1 \right) \,.
 \end{split}
\end{equation}

In order to obtain the locally non-geometric R-flux background we must dualise along the coordinates $\y$, $\z$ and $\w$. Because of the duality along the $\z$ direction, the $\z$ appearing in the metric \eqref{eq:TwistedTorusMetric} will become a dual coordinate $\tilde{x}_{24}$, in the $R$-flux background. This is analogous to what we saw for the string $R$-flux background. In fact, the resulting space is now fibred over the dual circle parameterised by $\tilde{x}_{24}$ making the background locally non-geometric. Furthermore, the $R$-flux background cannot be described using a metric and three-form $C_3$ since these would be ill-defined -- even on an ``extended space''. If one tried one would find
\begin{equation}
 \begin{split}
  ds_{11}^2 &= \left( 1 + N^2 \tilde{x}_{24}^2 \right)^{1/3} ds_7^2 + \left( 1 + N^2 \tilde{x}_{24}^2 \right)^{1/3} \pdz^2 \\
  & \quad + \left(1+N^2\tilde{x}_{24}^2\right)^{-2/3} \left( \pdx^2 + \pdy^2 + \pdw^2 \right) \,, \\
  C_3 &= \frac{N\tilde{x}_{24}}{1+N^2 \tilde{x}_{24}^2} \dx\wedge\dz\wedge\dw \,, \label{eq:MRFluxBF}
 \end{split}
\end{equation}
where we have included the external part of the metric to highlight the warping. This is ill-defined along the $\tilde{x}_{24}$ circle (albeit this is now in the dual space), where one would have to patch with a U-duality as $\tilde{x}_{24} \longrightarrow \tilde{x}_{24} + 1$.

Completely analogous to the string-theory case, we can instead use a trivector, which can be related by a field redefinition (or equivalently in exceptional generalised geometry or exceptional field theory by a $\mathrm{SO}(5)$ rotation, which changes the parameterisation of the generalised vielbein) \cite{Blair:2014zba}
\begin{equation}
 \begin{split}
  \hat{g}_{\alpha\beta} &= \left( 1 + V^2 \right)^{-1/3} \left[ \left(1+V^2\right) g_{\alpha\beta} - V_\alpha V_\beta \right] \,, \\
  \Omega^{\alpha\beta\gamma} &= \left( 1 + V^2 \right)^{-1} g^{\alpha\rho} g^{\beta\sigma} g^{\gamma\delta} C_{\rho\sigma\delta} \,, \\
  \hat{ds}_7^2 &= \left(1+V^2\right)^{-1/3} ds_7^2 \,. \label{eq:FieldRedef}
 \end{split}
\end{equation}
Here the indices $\alpha, \beta = 1, \ldots, 4$ and $V^\alpha = \frac{1}{3!|e|} \epsilon^{\alpha\beta\gamma\delta} C_{\beta\gamma\delta}$ with $\epsilon^{\alpha\beta\gamma\delta} = \pm 1$ the tensor density. Also we defined $V^2 = V^\alpha V^\beta g_{\alpha\beta}$.

Using these fields, the background is given by
\begin{equation}
 \hat{ds}_{11}^2 = ds_7^2 + \pdx^2 + \pdy^2 + \pdz^2 + \pdw^2 \,, \qquad \Omega^{134} = N \tilde{x}_{24} \,, \label{eq:MRFluxGF}
\end{equation}
and is clearly well-defined. The $R$-flux in M-theory is defined as \cite{Blair:2014zba}
\begin{equation}
 R^{\alpha,\beta\gamma\delta\rho} = 4 \hat{\partial}^{\alpha[\beta} \Omega^{\gamma\delta\rho]} \,,
\end{equation}
where $\hat{\partial}^{\alpha\beta} = \partial^{\alpha\beta} + \Omega^{\alpha\beta\gamma} \partial_\gamma$, with
\begin{equation}
 \partial^{\alpha\beta} = \frac{\partial}{\partial x_{\alpha\beta}} \,,
\end{equation}
the derivative with respect to the dual coordinates. The derivative $\hat{\partial}^{\alpha\beta}$ is an ``improved dual derivative'' \cite{Blair:2014zba} in order to obtain a spacetime tensor. We summarise this and other relevant results of \cite{Blair:2014zba} in Appendix \ref{s:A2}. Thus we find
\begin{equation}
 R^{4,1234} = N \,.
\end{equation}

In the case at hand, the field redefinition, or equivalently the $\mathrm{SO}(5)$ transformation, \eqref{eq:FieldRedef} is globally ill-defined since it relates an ill-defined frame \eqref{eq:MRFluxBF} to a well-defined one \eqref{eq:MRFluxGF}. A more careful analysis would observe that using the globally well-defined 1-forms of the twisted torus we can write down a globally well-defined generalised vielbein in the ``geometric frame''
\begin{equation}
 E_a{}^{\bar{a}} = e^{1/10} \begin{pmatrix}
  e_\alpha{}^{\bar{\alpha}}/\sqrt{e} & 0 \\ V^{\bar{\alpha}} & \sqrt{e}
 \end{pmatrix} \,, \label{eq:StdGeneralisedVielbein}
\end{equation}
where $\alpha = 1, \ldots, 4$ denotes spacetime indices, $\bar{\alpha}$ are the spacetime indices flattened by the spacetime vielbein $e_\alpha{}^{\bar{\alpha}}$. $e$ denotes the determinant fo the vielbein and $V^\alpha = \frac{1}{3!|e|} \epsilon^{\alpha\beta\gamma\delta} C_{\beta\gamma\delta}$ is the dualised 3-form, with $\epsilon^{\alpha\beta\gamma\delta} = \pm 1$ the alternating tensor density. We thus have for the twisted torus
\begin{equation}
 \left(E^{TT}\right)_a{}^{\bar{a}} = \begin{pmatrix}
  1 & 0 & 0 & 0 & 0 \\
  - N\z & 1 & 0 & 0 & 0 \\
  0 & 0 & 1 & 0 & 0 \\
  0 & 0 & 0 & 1 & 0 \\
  0 & 0 & 0 & 0 & 1
 \end{pmatrix} \,,
\end{equation}
where the $TT$ superscript stands for ``twisted torus''. Dualising to the $R$-flux background we have
\begin{equation}
 \left(E^R\right)_a{}^{\bar{a}} = \begin{pmatrix}
  1 & 0 & 0 & 0 & - N \tilde{x}_{24} \\
  0 & 1 & 0 & 0 & 0 \\
  0 & 0 & 1 & 0 & 0 \\
  0 & 0 & 0 & 1 & 0 \\
  0 & 0 & 0 & 0 & 1
 \end{pmatrix} \,,
\end{equation}
and we see that the parameterisation of the generalised vielbein \eqref{eq:StdGeneralisedVielbein} is not globally well-defined. However, there is a different parameterisation, also known as the ``non-geometric frame'' given by
\begin{equation}
 E_a{}^{\bar{a}} = e^{1/10} \begin{pmatrix}
  e_\alpha{}^{\bar{\alpha}}/\sqrt{e} & W_\alpha \\
  0 & \sqrt{e}
 \end{pmatrix} \,, \label{eq:NGGeneralisedVielbein}
\end{equation}
where $W_\alpha = \frac{|e|}{3!} \epsilon_{\alpha\beta\gamma\delta} \Omega^{\beta\gamma\delta}$ is dual to a trivector, the generalisation of the bivector $\beta^{ij}$ of string theory. We see immediately that this parameterisation is globally well-defined leading to \eqref{eq:MRFluxGF}.

\section{The non-associative algebras generated by octonions and their contractions}

\subsection{The Malcev algebra of imaginary octonions} \label{s:OctAssoc}
The imaginary units  $e_A$, (with $A = 1, \ldots, 7$) of octonions generate a simple Malcev algebra under the commutator product which is non-commutative and non-associative, and which we will often refer to as the algebra of imaginary octonions. For readers unfamiliar with the octonions, we summarise their relevant features in appendix \ref{s:AppOctonions}, including their multiplication rules in \ref{s:AppOctMult} following  \cite{Gunaydin:1973rs} and the definition of a Malcev algebra in \ref{s:AppMalcev}. For the applications considered in this paper we shall relable  three of the imaginary units as follows
\begin{equation}
 e_{(i+3)} = f_i \,, \qquad \textrm{for } i = 1, 2, 3 \,.
\end{equation}
In terms of $e_i, f_i$ and $e_7$ the multiplication table of octonions $\Oct$ takes the form
\begin{equation}
 \begin{split}
  e_i e_j &= - \delta_{ij} + \epsilon_{ijk} e_k\,, \\
  e_i f_j &= \delta_{ij} e_7 - \epsilon_{ijk} f_k\,, \\
  f_i f_j &= - \delta_{ij} - \epsilon_{ijk} e_k\,, \\
  e_7 e_i &= f_i\,, \qquad f_i e_7 = e_i \,,
 \end{split}
\end{equation}
with $e_A e_B = - e_B e_A$ whenever $A \neq B$. The commutators of the imaginary octonions  in this basis are given by 
\begin{equation}
 \begin{split}
  \left[ e_i, e_j \right] &= 2 \epsilon_{ijk} e_k \,, \qquad \left[ e_7, e_i \right] =  2 f_i \,, \\
  \left[ f_i, f_j \right] &= - 2 \epsilon_{ijk} e_k \,, \qquad \left[ e_7, f_i \right] = - 2 e_i \,, \\
  \left[ e_i, f_j \right] &= 2 \delta_{ij} e_7 - 2 \epsilon_{ijk} f_k \,. \label{eq:OctCommut}
 \end{split}
\end{equation}

We define the associator for any three imaginary octonions $e, f, g \in \Oct$ as
\begin{equation}
 \left[ e, f, g \right] = 2 (ef)g - 2 e(fg) \,. \label{eq:OctTriple}
\end{equation}
Note that with this convention the associator of the octonions is related to the Jacobiator by
\begin{equation}
 \left[ e, f, g \right] = \frac{1}{3} \left( \left[ \left[ e, f\right], g \right] + \left[ \left[g, e\right], f \right] + \left[ \left[ f, g\right], e \right] \right) = \frac13 \mathrm{Jac}\left(e,f,g\right) \,.
\end{equation}
The associator of the three imaginary units of any given quaternion subalgebra, such as that spanned by the $e_i$, ($i=1,2,3$) vanishes. The non-vanishing associators of the imaginary units of $\Oct$ in the above basis are as follows:
\begin{equation}
 \begin{split}
  \left[e_i, e_j, f_k\right] &= 4 \epsilon_{ijk} e_7 - 8 \delta_{k[i} f_{j]}  \,, \\
  \left[e_i, f_j, f_k \right] &= -8 \delta_{i[j} e_{k]} \,,\\
  \left[f_i,f_j, f_k\right] &= -4 \epsilon_{ijk} e_7 \,,\\
  \left[e_i, e_j, e_7\right] &= -4 \epsilon_{ijk} f_k \,,\\
  \left[e_i, f_j, e_7\right] &= -4 \epsilon_{ijk} e_k \,,\\
  \left[f_i, f_j, e_7\right] &= 4\epsilon_{ijk} f_k \,. \label{eq:OctAssoc}
 \end{split}
\end{equation}

\subsection{Contraction of the Malcev algebra of octonions to the string $R$-flux algebra} \label{s:Contraction}
We will now show that the string $R$-flux algebra can be obtained by contracting the octonionic Malcev algebra given in equations \eqref{eq:OctCommut} and \eqref{eq:OctAssoc}. Recall that the coordinates and momenta of strings in the $R$-flux background, $x^i$ and $p_i$, have been shown to form a non-associative algebra \cite{Lust:2010iy,Blumenhagen:2010hj,Blumenhagen:2011ph,Lust:2012fp,Blumenhagen:2013aia}\footnote{This algebra also shows up in the other T-duality frames with $H$-, $f$- and $Q$-flux respectively \cite{Blumenhagen:2011ph} as it is clear also from the view point of double field theory \cite{Bakas:2013jwa,Blumenhagen:2013zpa}, but we restrict the discussion to the $R$-flux background.} whose only non-vanishing relations are
\begin{equation}
 \left[ x^i, x^j, x^k \right] = R^{ijk} \,, \qquad \left[ x^i, x^j \right] = i R^{ijk} p_k \,, \qquad \left[ x^i, p_j \right] = i \delta^i_j \,.\label{onestring}
\end{equation}
The triple-bracket here is the associator defined in terms of commutators as
\begin{equation}
 \left[ x^1, x^2, x^3 \right] = \frac13 \left( \left[ \left[ x^1, x^2\right], x^3 \right] + \left[ \left[x^3, x^1\right], x^2 \right] + \left[ \left[ x^2, x^3\right], x^1 \right] \right) \,.
\end{equation}
For the moment we have set the constants $\hbar$ and $l_s$, which appeared in equations \eqref{commclosed}-\eqref{asso}, to unity but we will comment on them later on. For the parabolic $R$-flux model on $T^3$, the $R$-flux is just proportional to the epsilon tensor: $R^{ijk} = N \epsilon^{ijk}$.

In order to recover this algebra from the Malcev algebra generated by imaginary octonions we define the three string momenta $p_i$ as the contraction of the three imaginary units $e_i$ and the three string coordinates $x^i$ as a contraction of the three imaginary unites $f_i$ as follows
\begin{equation}
 p_i = - i \lambda \frac{1}{2} e_i  \, ,\qquad x^i = i \lambda^{1/2} \frac{\sqrt{N}}{2} f_i \,, \label{cona}
 \end{equation}
where in a moment we will take the limit $\lambda\rightarrow 0$. Furthermore, we define the contraction of the seventh imaginary unit as
\begin{equation}
 I = i  \lambda^{3/2} \frac{\sqrt{N}}{2} e_7 \, .\label{conb}
\end{equation}
Now the contraction of the octonionic algebra is defined by taking the limit $\lambda\rightarrow 0$. In this way we obtain
\begin{equation}
  \left[f_i , f_j\right] = - 2 \epsilon_{ijk} e_k\quad \Longrightarrow \quad [ x^i ,x^j ] = i N \epsilon^{ijk} p_k \,,
\end{equation}
and
\begin{equation}
 \left[e_i , e_j\right] = 2 \epsilon_{ijk} e_k \quad \Longrightarrow\quad [ p_i ,p_j ] = 0 \,,
 \end{equation}
which shows that the quaternionic subalgebra of the momenta becomes completely commuting after contraction. Finally due to the contraction of the seventh imaginary unit one derives
\begin{equation}
 \begin{split}
  \left[x^i , p_j\right] &= \lim_{\lambda \rightarrow 0} \left( \frac{\lambda^{3/2}\sqrt{N}}{4} \left[ f_i , e_j \right]\right)  =  \lim_{\lambda \rightarrow 0} (\frac{\lambda^{3/2}\sqrt{N}}{2} ) \left[ -\delta^{i}_{j} e_7  + \epsilon^{i}{}_{jk} f_k \right] \\
  &= i \delta^{i}_{j} I +  \lim_{\lambda \rightarrow 0} \epsilon^{i}{}_{jk} ( \lambda x_k)  = i \delta^{i}_{j} I \,, \label{eq:xpstringJ}
 \end{split}
\end{equation}
and
\begin{equation}
 \left[ x_i , I\right] =0 = \left[p_i , I \right] \,.
\end{equation}
$I$ is thus a central element of the contracted algebra and can be taken to be the identity operator. 
Finally, the only non-vanishing associator after contraction is that of three coordinates
\begin{equation}
 \left[f_i,f_j, f_k\right] = -4 \epsilon_{ijk} e_7 \quad \Longrightarrow \quad \left[x^i, x^j, x^k \right] = N \epsilon^{ijk} I \,.
\end{equation}
Now by replacing $N\epsilon^{ijk}$ with $R^{ijk}$ we see that the contracted algebra is indeed the string theory $R$-flux algebra \eqref{onestring}.
\subsection{The M-theory $R$-flux algebra} \label{s:MNonAssoc}
Given that the non-associative string $R$-flux algebra is recovered by a contraction of the Malcev algebra of imaginary octonions, it is natural to ask what, if any, role the full uncontracted algebra of imaginary octonions plays. Here we propose that it gives the uplift of the string $R$-flux algebra to M-theory. That is, we conjecture that M-theory with locally non-geometric flux also has a non-associative structure which for four-dimensional backgrounds is given by the Malcev algebra of imaginary octonions as we will make precise in section \ref{s:MAlgebra}.

First we wish to address how one can identify the seven imaginary octonions $e_i$, $f_i$ and $e_7$ with the phase space of membranes moving in a four-dimensional M-theory background. Naively, this is not possible as one would expect the phase space to be eight-dimensional. However, we will now show that the phase space for locally non-geometric backgrounds is indeed seven-dimensional, providing evidence for our conjecture.

\subsubsection{A lack of momentum}
Let us consider the toy model discussed in section \ref{s:MToy} where a locally non-geometric M-theory background is obtained by U-duality from a twisted torus. We will show that in this case the locally non-geometric background is ``missing'' a momentum mode, and this allows us to identify the seven imaginary octonions with the phase space variables, consisting of four coordinates $X^\alpha$ and three momenta $P_i$.

Under U-duality momentum modes are exchanged with wrapping numbers, which are classified by homology. As a result, the homology of the twisted torus also determines the possible momenta of the $R$-flux background. A ``missing'' cycle in the homology -- compared to a toroidal background -- implies that the dual background is missing a momentum mode, compared to naive expectations. This phenomenon occurs for the M-theory twisted torus and its dual locally non-geometric background as we will now explain.

Under U-duality along the $\y$, $\z$, $\w$ directions, the following wrapping numbers of the twisted torus, where they exist, become momenta of the $R$-flux background
\begin{equation}
 W^{23} \longrightarrow P_4 \,, \qquad W^{42} \longrightarrow P_3 \,, \qquad W^{34} \longrightarrow P_2 \,.
\end{equation}
The twisted background \eqref{eq:TwistedTorusMetric} is just $ \tilde{T}^3\times S^1$, where $S^1$ is the $\w$-circle and $\tilde{T}^3$ is the twisted torus parameterised by the $(\x,\y,\z)$ coordinates. Therefore the 2-cycles corresponding to $W^{42}$ and $W^{34}$ are the tori whose 1-cycles are the $\w$-cycle and $\y$-/$\z$-cycles, respectively, of the $\tilde{T}^3$. On the other hand, the cycle of $W^{23}$ corresponds to the $(\y\z)$ 2-cycle of $\tilde{T}^3$.

However, for there to be a wrapping number for these cycles, they need to be homologically non-trivial. We will now see that this is the case for the $(\w\y)$- and $(\z\w)$-cycles but not for the $(\y\z)$-cycles, and so there is no $W^{23}$ wrapping number. Let us begin by studying the deRham cohomology of $\tilde{T}^3$. Recall from \eqref{eq:TwistedTorus1Forms} that the globally well-defined 1-forms on $\tilde{T}^3$ are
\begin{equation}
 \eta^1 = \dx - N \z \dy \,, \qquad \eta^2 = \dy \,, \qquad \eta^3 = \dz \,.
\end{equation}
None of these are exact but not all three are closed since
\begin{equation}
 d\eta^2 = d\eta^3 = 0 \,, \qquad d\eta^1 = N \dy \wedge \dz \neq 0 \,.
\end{equation}
Thus $H^1(\tilde{T}^3,\mathbb{R}) = \mathbb{R}^2$ and is generated by the $\y$- and $\z$-cycles so there are wrapping numbers $W^{42}$ and $W^{34}$.

For the second cohomology we could just apply Poincar\'{e} duality to see that there is no $(\y\z)$-cycle and so $H^2(\tilde{T}^3,\mathbb{R}) = \mathbb{R}^2$. Equivalently, we can consider the three globally-defined two-forms
\begin{equation}
 \begin{split}
  \kappa^{12}& = \eta^1\wedge\eta^2 = \dx \wedge \dy \,, \\
  \kappa^{13} &= \eta^1\wedge\eta^3 = \dx \wedge \dz - N\z\dy\wedge\dz \,, \\
  \kappa^{23} &= \eta^2\wedge\eta^4 = \dy \wedge \dz \,.
 \end{split}
\end{equation}
While these are all closed, $\kappa^{23} = \frac{1}{N} d\eta^1$ is exact. So as expected there are only two elements of the second deRham cohomology, generated by $\kappa^{12}$ and $\kappa^{13}$. There is no $(\y\z)$-cycle and consequently no wrapping number $W^{23}$.

The above argument is actually naive because the wrapping numbers take values in the integer homology groups and these can have torsion parts. In fact $H_1(\tilde{T}^3,\mathbb{Z}) = \mathbb{Z}^2 \oplus \mathbb{Z}_N$, as can be seen from Hurewicz's theorem for path-connected spaces $X$, which says
\begin{equation}
 H_1(X,\mathbb{Z}) \simeq \pi_1^{ab}(X) \,,
\end{equation}
i.e. the integral first homology group is the abelianisation of the fundamental group. Here this is easily seen to be $\pi_1^{ab}(\tilde{T}^3) = \mathbb{Z}^2 \oplus \mathbb{Z}_N$, see for example \cite{Kachru:2002sk}. The $\mathbb{Z}^2$ in this case is generated by the $\y$ and $\z$-cycles so that we still have wrapping numbers $W^{42}$ and $W^{34}$, which are now quantised as expected.\footnote{The torsion part corresponds to the $\x$-direction and plays a role in string theory compactified on the twisted torus, as discussed in \cite{Kachru:2002sk}. There the finite order of windings along the $\x$-direction, taking values in $\mathbb{Z}_N$, are dual to momenta in the $T^3$ background with $H$-flux which are only conserved modulo $N$ due to the $B$-field.}

However, $H_2(\tilde{T}^3,\mathbb{Z})$ cannot have torsion since $\tilde{T}^3$ is oriented. This follows from the fact that for any $n$-dimensional oriented topological space $X$, $H_{n-1}(X,\mathbb{Z})$ must be torsion-free. Therefore, the naive deRham cohomology computation is correct in this case. As a result there is no momentum $P_4$ along the $\w$ directions in the dual $R$-flux background and this is a key insight to allow us to identify the non-associative $R$-flux algebra with that of octonions discussed in section \ref{s:OctAssoc}.

The fact that there is no momentum in the $X^4$ direction means that in the appropriate string theory picture there cannot be D0-branes. Indeed, this agrees with \cite{Wecht:2007wu} where it is argued that there are no D0-branes in the string $R$-flux model because D3-branes cannot wrap a $T^3$ with $H$-flux due to the Freed-Witten anomaly. By applying three T-dualities to the $T^3$ with $H$-flux one obtains the string $R$-flux background and, conversely, D0-branes in the $R$-flux backgrounds would be dual to D3-branes on $T^3$ with $H$-flux which do not exist. Hence the $R$-flux background cannot support D0-branes. Our M-theory consideration lead to the same conclusion if we take the $X^4$ direction to be a vanishing circle.

Motivated by our findings for this toy model we postulate that for a general background with $R$-flux $R^{\alpha,\beta\gamma\delta\rho}$, the momenta satisfy the constraints
\begin{equation}
 P_\alpha R^{\alpha,\beta\gamma\delta\rho} = 0 \,, \label{eq:MomentumConstraint}
\end{equation}
which implies that some momentum modes are missing. Indeed, one could more generally consider dualising a $\mathrm{U}(1)$-fibration with non-vanishing 1st Chern Class in such a way as to obtain a locally non-geometric background. Because the 1st Chern Class is trivial in the total space of the $\mathrm{U}(1)$-fibration we again expect a missing momentum mode in the dual $R$-flux background.

As discussed above, this lack of momentum implies there can be no D0-branes in the string theory limit, and this in turn is related to the Freed-Witten anomaly by duality. This suggests that the M-theory constraint \eqref{eq:MomentumConstraint} may be related to a membrane anomaly cancellation condition and we would hope that a duality-invariant study of membranes leads to a constraint of the form of \eqref{eq:MomentumConstraint}.

\subsubsection{The M-theory $R$-flux algebra} \label{s:MAlgebra}
As we have just shown the phase space of the locally non-geometric background is seven-dimensional. In light of the contraction to the string algebra, discussed in \ref{s:Contraction}, we now conjecture that for the parabolic model the coordinates and momenta are given in terms of the imaginary octonions by
\begin{equation}
 X^i = \frac12 i \sqrt{N} f_i \,, \qquad X^4 =  \frac12 i \sqrt{N} e_7 \,, \qquad P_i = - \frac12 i e_i \,, \label{eq:XPnodimensions}
\end{equation}
where $i = 1, \ldots, 3$. Here the non-vanishing $R$-flux is given by (see section \ref{s:MToy} and appendix \ref{s:AppRflux} )
\begin{equation}
 R^{4,\alpha\beta\gamma\delta} = N \epsilon^{\alpha\beta\gamma\delta} \,,
\end{equation}
 so that the above parameterisation of the phase space \eqref{eq:XPnodimensions} satisfies the constraint \eqref{eq:MomentumConstraint}. We will also write the four coordinates as $X^{\alpha} = \left( X^i, X^4 \right)$ with $\alpha = 1, \ldots, 4$.

These variables now generate a non-commutative and non-associative algebra with commutators
\begin{equation}
 \begin{split}
  \left[ X^i, X^j \right] &= i N \epsilon^{ijk} P_k \,, \qquad \left[X^4, X^i \right] = i N P^i \,, \\
  \left[ P_i, P_j \right] &= - i \epsilon_{ijk} P^k \,, \qquad \left[ P_i, X^4 \right] = i  X_i \,, \\
  \left[ X^i, P_j \right] &=i \delta^i_j X^4 + i \epsilon^{i}{}_{jk} X^k \,,
 \end{split}
\end{equation}
and non-vanishing associators
\begin{equation}
 \begin{split}
  \left[ X^\alpha, X^\beta, X^\gamma \right] &= N \epsilon^{\alpha\beta\gamma\delta} X_\delta \,, \\
  \left[ P_i, X_j, X_k \right] &= 2 N \delta_{i[j} P_{k]} \,, \\
  \left[ P_i, X_j, X_4 \right] &= N \epsilon_{ijk} P^k \,, \\
    \left[ P_i, P_j, X_\alpha \right] &= - \epsilon_{ij\alpha\beta} X_\beta + 2 \delta_{\alpha[i} X_{j]} \,. \label{eq:MAlgebraNoUnits}
 \end{split}
\end{equation}
The $P_i$ define a Lie subalgebra with vanishing Jacobiators. Throughout we raise and lower indices with $\delta_{\alpha\beta}$. This makes sense because the $\SOf$ subgroup of the $\textrm{G}_2$ automorphism group of octonions preserves the above split into momenta and coordinates. We rewrite the algebra in a manifestly $\SOf$-invariant manner in appendix \ref{s:AppSO4}.

We now recognise the combinations $N \epsilon^{\alpha\beta\gamma\delta} = R^{4,\alpha\beta\gamma\delta}$ as the $R$-flux tensor and thus we write the algebra as
\begin{equation}
 \begin{split}
  \left[ X^i, X^j \right] &= i R^{4,ijk4} P_k \,, \qquad \left[X^4, X^i \right] = i R^{4,1234} P^i \,, \\
  \left[ P_i, P_j \right] &= -i \epsilon_{ijk} P^k \,, \qquad \left[ P_i, X^4 \right] = i X_i \,, \\
  \left[ X^i, P_j \right] &= i \delta^i_j X^4 + i \epsilon^{i}{}_{jk} X^k \,, \\
  \left[ X^\alpha, X^\beta, X^\gamma \right] &= R^{4,\alpha\beta\gamma\delta} X_\delta \,, \\
  \left[ P_i, X^j, X^k \right] &= 2 R^{4,1234} \delta_{i}^{[j} P^{k]} \,, \\
  \left[ P^i, X^j, X^4 \right] &= R^{4,ijk4} P_k \,, \\
  \left[ P_i, P_j, X_\alpha \right] &= - \epsilon_{ij\alpha\beta} X_\beta + 2 \delta_{\alpha[i} X_{j]} \,. \label{eq:MAlgebraR}
 \end{split}
\end{equation}
We see that the algebra is compatible with the form of the $R$-flux tensor as given by exceptional field theory \cite{Blair:2014zba}. In particular the relationship
\begin{equation}
 \left[ X^\alpha, X^\beta, X^\gamma \right] = R^{4,\alpha\beta\gamma\delta} X_\delta \,,
\end{equation}
is a natural generalisation of \eqref{onestring}.

The algebra \eqref{eq:MAlgebraR} given in the form above is expressed in terms of preferred coordinates, where we have broken the diffeomorphism-invariance by solving the missing-momentum constraint \eqref{eq:MomentumConstraint} explicitly. There is no three-dimensional representation of $\SL{4}$ and hence solving this constraint so that only three momenta survive necessarily leads to an algebra which is not $\SL{4}$-invariant.

Indeed, the above algebra is invariant under $\SOf$, as further discussed in \ref{s:AppSO4}, which does have a three-dimensional representation. However, this $\SOf$ group does not act on the spacetime indices in a consistent way since it treats $X^\alpha$ as a vector while leaving $R^{4,\alpha\beta\gamma\delta} = \epsilon^{\alpha\beta\gamma\delta}$ invariant. It is not clear to us whether this $\SOf$ symmetry is physical. One should also note that the algebra \eqref{eq:MAlgebraR} is not invariant under $\SL{3}$ either but that this symmetry is restored upon contraction as we have seen in section \ref{s:Contraction}.

We have used dualities to justify our conjectured constraint \eqref{eq:MomentumConstraint} which produces a seven dimensional phase space. However, it would be nice to understand the constraint in an appropriate mathematical framework, given its unusual feature of producing an \emph{odd}-dimensional phase space. For example, one may wonder whether there is a way of implementing the constraint \eqref{eq:MomentumConstraint} in a covariant manner, say by some form of ``Nambu-Dirac bracket'' which implements the phase space constraint while still violating the Jacobi identity. This presumably should allow us to rewrite \eqref{eq:MAlgebraR} in a manifestly diffeomorphism-invariant manner where indices are not raised or lowered by $\delta_{\alpha\beta}$. It would then be interesting to see how such a Nambu-Dirac bracket reduces the phase-space dimension by an odd number and what the corresponding notion of first- and second-class constraints are.

We also note that there are two kinds of modifications of the phase space algebra compared to the flat-space case
\begin{equation}
 \left[ X^\alpha, X^\beta \right] = \left[ P_\alpha, P_\beta \right] = 0 \,, \qquad \left[ X^\alpha, P_\beta \right] = i \delta^\alpha_\beta \,.
\end{equation}
One is proportional to $R^{4,\alpha\beta\gamma\delta}$ and increases as $N$ increases, while the other modification is independent of $N$ (though of course requires $N \neq 0$). While the first modification is linear in the flux and survives the contraction process, the second modification is due to the fact that there is a missing momentum mode and will not survive the contraction process. Finally, let us note that both of these modifications are necessary in order for the algebra to be invariant under $\SOf$.

\subsubsection{The dimensionful M-theory algebra} \label{s:DimAlgebra}
We have seen that the string $R$-flux algebra can be obtained as a contraction of the algebra of imaginary octonions and we proposed that the uncontracted algebra of imaginary octonions is the uplift of the $R$-flux algebra to M-theory. Thus we see that $\lambda\rightarrow 0$ plays the role of going from M-theory to the type IIA string by taking the limit of weak string coupling $g_s\rightarrow 0$ or equivalently the  limit of vanishing radius of the 11th direction $R_{11}\rightarrow0$. Hence it is natural to identify the contraction parameter $\lambda$ with the string coupling constant, with $g_s \rightarrow 0$ as $\lambda \rightarrow 0$. However, there may not be a simple polynomial relationship between $\lambda$ and $g_s$. Instead $\lambda$ may approach a finite value as $g_s \rightarrow \infty$.

Now we can finally go back to uncontracted algebra \eqref{eq:MAlgebraR} and introduce $\lambda$, as follows from the contraction \eqref{cona} and \eqref{conb}. We conjecture $\lambda$ to be related to the string coupling constant $g_s$ as discussed above, with $\lambda \rightarrow 0$ as $g_s \rightarrow 0$. Furthermore we also want to re-introduce $\hbar$ and string length $l_s$ at their relevant positions. This involves defining the positions and momenta in terms of the imaginary octonions as
\begin{equation}
 X^i = \frac12 i \sqrt{N} l_s^{3/2} \lambda^{1/2} f_i \,, \qquad X^4 = \frac12 i \sqrt{N} l_s^{3/2} \lambda^{3/2} e_7 \,, \qquad P^i = - \frac12 i \hbar \lambda e_i \,.
\end{equation}

Then the full non-associative M-theory algebra takes the following final form:
\begin{equation}
 \begin{split}
  \left[ P_i, P_j \right] &= - i \lambda \hbar \epsilon_{ijk} P^k \,, \qquad \left[ X^4, P_i \right] = i \lambda^2 \hbar X_i \,, \\
  \left[ X^i, X^j \right] &= \frac{il_s^3}{\hbar} R^{4,ijk4} P_k \,, \qquad \left[ X^4, X^i \right] = \frac{i \lambda l_s^3}{\hbar} R^{4,1234} P^i \,, \\
  \left[ X^i, P_j \right] &= i \hbar \delta^{i}_{j} X^4 + i \lambda \hbar \epsilon^{i}{}_{jk} X^k \,, \\
  \left[ X^i, X^j, X^k \right] &= l_s^3 R^{4,ijk4} X_4 \,, \\
  \left[ X^i, X^j, X^4 \right] &= - \lambda^2 l_s^4 R^{4,ijk4} X_k \,, \\
  \left[ P_i, X^j, X^k \right] &= 2 \lambda l_s^3 R^{4,1234} \delta_{i}^{[j} P^{k]} \,, \\
  \left[ P^i, X^j, X^4 \right] &= \lambda^2 l_s^3 R^{4,ijk4} P_k \,, \\
  \left[ P_i, P_j, X_k \right] &= - \lambda^2 \hbar^2 \epsilon_{ijk} X^4 + 2 \lambda \hbar^2 \delta_{k[i} X_{j]} \,, \\
  \left[ P_i, P_j, X_4 \right] &= \lambda^3 \hbar^2 \epsilon_{ijk} X_k \,, \\
  \left[ P_i, P_j, P_k \right] &= 0 \,. \label{eq:FullMAlgebra}
 \end{split}
\end{equation}
As shown before, in the limit $\lambda \rightarrow 0$ (i.e. $g_s\rightarrow 0$) this M-theory algebra correctly reduces to the string $R$-flux algebra. While we do not know the precise relationship between $\lambda$ and $g_s$, we speculate that $\lambda$ has a finite value as $g_s \rightarrow \infty$ so that the above algebra could still be made sense of in the strong-coupling regime.\footnote{One may wonder how one should define non-geometric fluxes in a non-compact setting, where the usual picture of a U-duality valued monodromy fails. We note that one could consider the duality with the twisted torus as a definition. A careful analysis shows that the value of $\Omega^{ijk}$ remains finite in the limit that the volume of the dual twisted torus is taken to vanish. $\Omega^{ijk}$ is in fact independent of the dual volume.}

\section{Discussion} \label{s:Discussion}
In this paper, we have shown how the non-associative phase space algebra for string $R$-flux backgrounds can be obtained by a contraction of the imaginary octonions. We further proposed that the full uncontracted algebra of imaginary octonions provides an uplift of the $R$-flux algebra to M-theory. We thus conjecture that locally non-geometric M-theory backgrounds also exhibit non-associativity amongst their coordinates.

We then showed using a four-dimensional background which is dual to a twisted torus and serves as a toy-model for a locally non-geometric M-theory background that there is a missing momentum mode. As a result, the phase space is indeed seven-dimensional and substantiates our non-associative proposal for M-theory. The algebra, as we have written it, is not invariant under diffeomorphisms, because of the lack of the momentum mode.

A crucial aspect of our proposal is the missing momentum mode for the locally non-geometric background which we proposed can in general be implemented as a constraint on phase space of the form
\begin{equation}
 R^{\alpha,\beta\gamma\delta\rho} P_\alpha = 0 \,.
\end{equation}
This arises because the dual background, the twisted torus, is \emph{not} a torus. The missing momentum is also related by duality to the Freed-Witten anomaly. We hope that this constraint can be recovered in this way from a duality-invariant study of membranes, although the M-theory lift of the Freed-Witten anomaly is very subtle \cite{Diaconescu:2000wy}.

It would also be interesting to understand if the missing momentum constraint can be imposed in a covariant manner by some form of ``Nambu-Dirac bracket'', and how this leads to the unusual feature of producing an \emph{odd}-dimensional phase space. We also hope that this would lead to a diffeomorphism-invariant form of the non-associative algebra we propose \eqref{eq:FullMAlgebra}.

A natural question arising out of this work is how the algebra given above generalises to higher dimensions. In this context we should note that the 27-dimensional auxiliary space of double field theory relevant to supergravity in five-dimensions can be identified with the generalised space-time coordinatised by the exceptional Jordan algebra over split octonions. The generalised diffeomorphism group of this generalised space-time is the exceptional group $E_{6(6)}$ which is the U-duality group of five-dimensional maximal supergravity\footnote{The concept of generalised space-times coordinatised by Jordan algebras was introduced in the early days of space-time supersymmetry \cite{Gunaydin:1975mp}. For the explicit construction of the symmetry groups defined over the split exceptional Jordan algebra we refer to \cite{Gunaydin:2000xr,Gunaydin:2009zza} and the references therein.}. Due to intrinsic non-associativity of the exceptional Jordan algebra we expect the corresponding non-geometric phases of the uplift to M-theory to describe the extensions of the results of this paper.  

Finally, we believe that the missing momentum we observe also clarifies the meaning of the doubled and extended coordinates in double and exceptional field theory, especially when the background is not a torus. In the original work of Hull and Zwiebach \cite{Hull:2009mi} the doubled coordinate space was understood as dual to the momenta and winding modes of strings propagating on a torus. With this interpretation it is not immediately clear what the ``extra'' coordinates mean when the background is topologically not a torus.

Here we propose that the coordinates are not necessarily linked to the topology of the background. Instead one can, in double field theory, view them as constants of integration of independent left/right-movers of the worldsheet CFT. We also expect that a similar interpretation of extended coordinates of exceptional field theory will arise from an appropriate quantum theory.

With this interpretation there is a dual background parameterised by the extra coordinates, and as we have explained, if the original background is not a torus, then its dual will in general have missing momentum modes. A similar effect has also been observed when dualising along an isometry which has singular points \cite{Rocek:1991ps}. There, it was argued that the dual background to flat space is a singular throat which has no normalisable momentum modes, but does have winding modes. The effect we observe here is similar in spirit, but occurs for non-singular, albeit locally non-geometric backgrounds.

\section*{Acknowledgements}
We would like to thank Daniel Waldram for helpful correspondence, Erik Plauschinn for discussions and Oleksandr Pavlyk for his help with figure 1. MG would also like to thank the hospitality of Ludwig-Maximilians-Universit\"at M\"unchen and NORDITA, Stockholm for their hospitality where part of this work was performed. The research of MG was supported in part under DOE Grant No: de-sc0010534. The work of EM and DL is supported by the ERC Advanced Grant ``Strings and Gravity" (Grant No. 32004).

\appendix

\section{Non-geometric fluxes} \label{s:AppRflux}
In this appendix we wish to summarise some of the relevant features of non-geometric backgrounds, mostly as discussed in \cite{Grana:2008yw,Andriot:2011uh,Andriot:2012an,Andriot:2012wx,Blumenhagen:2012nt,Blumenhagen:2013aia} for string theory and \cite{Blair:2014zba} for M-theory.

\subsection{$R$-flux in string theory} \label{s:A1}
Locally non-geometric backgrounds in string theory can be characterised by a tensor $R^{ijk}$, which measures the so-called $R$-flux. It can also be seen as a component of the embedding tensor of gauged supergravities \cite{Aldazabal:2011nj,Aldazabal:2013mya,Berman:2013uda}. In order to define it as a spacetime tensor, it is important to understand how the supergravity fields of the ``non-geometric frame'', $g_{ij}$ and $\beta^{ij}$ transform under spacetime diffeomorphisms \cite{Andriot:2012an,Andriot:2012wx}. We also wish to highlight that the non-geometric frame can be understood in terms of Lie algebroids as discussed in \cite{Blumenhagen:2012nt,Blumenhagen:2013aia}.

\paragraph{Transformation under spacetime diffeomorphisms}
Recall that the generalised metric in ``non-geometric frame'' (i.e. in terms of $g_{ij}$ and $\beta^{ij}$) takes the form \cite{Grana:2008yw,Andriot:2011uh}.
\begin{equation}
 M_{IJ} = \begin{pmatrix}
  g_{ij} & g_{ik} \beta^{kj} \\ - \beta^{ik} g_{kj} & g^{ij} - \beta^{ik} g_{kl} \beta^{lj}
 \end{pmatrix} \,,
\end{equation}
where $I = 1, \ldots, 2D$ denote $\mathrm{O}(D,D)$ indices. Now consider the generalised Lie derivative \cite{Hull:2009mi}
\begin{equation}
 \gL_U V^I = U^J \partial_J V^I - V^J \partial_J U^I + \eta^{IK} \eta_{JL} V^J \partial_K U^L \,,
\end{equation}
which is the $\mathrm{O}(D,D)$ extension of the standard Lie derivative. $U$ and $V$ are $\mathrm{O}(D,D)$ vectors, e.g. $U^I = \left( \xi^i, \tilde{\xi}_i \right)$ consists of a vector piece and a 1-form piece. If we act with the generalised Lie derivative on the generalised metric $M_{IJ}$ we can read off the transformation rules for $g_{ij}$ and $\beta^{ij}$ with respect to the symmetries generated by a vector, $\xi^i$, and a 1-form, $\tilde{\xi}_i$. The symmetries generated by $\xi^i$ are spacetime diffeomorphisms while those generated by $\tilde{\xi}_i$ are a sort of gauge symmetry.  We will completely ignore the latter and focus on just the spacetime diffeomorphisms generated by $\xi^i$. We find \cite{Andriot:2011uh,Andriot:2012wx} that $g_{ij}$ and $\beta^{ij}$ transform as
\begin{equation}
 \begin{split}
  \delta_\xi g_{ij} &= L_\xi g_{ij} = \xi^k \partial_k g_{ij} + 2 g_{k(i} \partial_{j)} \xi^k \,, \\
  \delta_\xi \beta^{ij} &= L_\xi \beta^{ij} - 2 \tilde{\partial}^{[i} \xi^{j]} \,, \label{eq:BetaTranformation}
 \end{split}
\end{equation}
where
\begin{equation}
 L_\xi \beta^{ij} = \xi^k \partial_k \beta^{ij} - 2 \beta^{k[j} \partial_k \xi^{i]} \,,
\end{equation}
is the tensorial action of the spacetime Lie derivative on $\beta^{ij}$.

Before moving on, we should highlight that the algebra of generalised diffeomorphisms closes only subject to the ``section condition'' or ``strong constraint'' of double field theory \cite{Hull:2009mi}, which says that for all fields $f$, $g$
\begin{equation}
 \eta^{IJ} \partial_I f \partial_J g = \partial_i f \tpartial^i g + \tpartial^i f \partial_i g = 0 \,, \qquad \eta^{IJ} \partial_I \partial_J f = 0 \,.
\end{equation}
Here
\begin{equation}
 \eta^{IJ} = \begin{pmatrix}
  0 & \delta_i{}^j \\ \delta^i{}_j & 0
 \end{pmatrix} \,,
\end{equation}
is the flat $\mathrm{O}(D,D)$ metric.

\paragraph{Improved dual derivative}
Looking at $\tilde{\partial}^i$ one can see that it is not a good derivative. This means in particular that
\begin{equation}
 \delta_\xi \tilde{\partial}^i \phi \neq L_\xi \tilde{\partial}^i \phi \,.
\end{equation}
Instead one finds
\begin{equation}
 \delta_\xi \tilde{\partial}^i \phi = L_\xi \tilde{\partial}^i \phi + \tpartial^i \xi^j \partial_j \phi + \tpartial^j \phi \partial_j \xi^i \,.
\end{equation}
However, one can define an improved derivative using $\beta^{ij}$
\begin{equation}
 \hpartial^i \phi = \tpartial^i \phi + \beta^{ij} \partial_j \phi \,, \label{eq:ImprovedWinding}
\end{equation}
and this transforms as
\begin{equation}
 \delta_\xi \hpartial^i \phi = L_\xi \hpartial^i \phi + \tpartial^j \xi^i \partial_j \phi + \partial_j \xi^i \tpartial^j \phi \,. \label{eq:ImprovedWindingTransformation}
\end{equation}
The anomalous terms vanish by the ``strong constraint'' (or section condition). Note something nice about the improved winding derivative is that it also satisfies the strong constraint
\begin{equation}
 \partial_i f \hpartial^i g + \hpartial^i f \partial_i g = 0 \,.
\end{equation}
However, we will not make use of this here.

\paragraph{R-flux as a spacetime tensor}
Now consider the following derivatives of $\beta^{ij}$, called the R-flux tensor,
\begin{equation}
 R^{ijk} = 3 \hpartial^{[i} \beta^{jk]} \,. \label{eq:StringRflux}
\end{equation}
This transforms as a spacetime tensor as can easily be seen by computing its variation under spacetime diffeomorphisms. Begin with

\begin{equation}
 \begin{split}
  \delta_\xi \hpartial^i \beta^{jk} &= \xi^k \partial_k \hpartial^{i} \beta^{jk} - \hpartial^l \beta^{jk} \partial_l \xi^i - 2 \hpartial^i \beta^{l[k} \partial_l \xi^{j]} + 2 \beta^{l[k} \hpartial^{|i|} \partial_l \xi^{j]} - 2 \hpartial^i \tpartial^{[j} \xi^{k]} \\
  &= L_\xi \tpartial^i \beta^{jk} - 2 \beta^{l[k} \partial_l \tpartial^{|i|} \xi^{j]} - 2 \beta^{im} \beta^{l[k} \partial_m \partial_l \xi^{j]} - 2 \tpartial^i \tpartial^j \xi^k + 2 \beta^{li} \partial_l \tpartial^{[j} \xi^{k]} \,,
 \end{split}
\end{equation}
where in going to the second line we have identified the first tree terms of the first line as the standard Lie derivative of $\tpartial^{i} \beta^{jk}$. If we now antisymmetrise over $i, j, k$ we the second and final term will cancel, while the third and fourth term will both vanish because of symmetry of the derivatives $\partial_{[i} \partial_{j]} = 0$ and $\tpartial^{[i} \tpartial^{j]} = 0$. Hence we find
\begin{equation}
 \delta_\xi R^{ijk} = L_\xi R^{ijk} \,.
\end{equation}
Although it looks as if we did not have to use the section condition, we did use it in order for the index on the dual derivative $\hpartial^i$ to transform as a vector, exactly as in \eqref{eq:ImprovedWindingTransformation}.
\subsection{Locally non-geometric $R$-flux in M-theory} \label{s:A2}
Here we review and summarise the relevant results of \cite{Blair:2014zba}.

\paragraph{Transformations and improved winding derivative}
The M-theory generalisation is very similar and we will not give details for the analogous steps. The detailed calculation is given in \cite{Blair:2014zba}. In the ``non-geometric frame'' where we have fields $g_{\alpha\beta}$ and $\Omega^{\alpha\beta\gamma}$ we have
\begin{equation}
 \delta_\xi g_{\alpha\beta} = L_\xi g_{\alpha\beta} \,, \qquad \delta_\xi \Omega^{\alpha\beta\gamma} = L_\xi \Omega^{\alpha\beta\gamma} - 3 \tpartial^{[\alpha\beta} \xi^{\gamma]} \,,
\end{equation}
where $L_\xi$ denotes the usual spacetime Lie derivative and $\tpartial^{\alpha\beta}$ denotes the dual derivative. Completely analogously to the string theory case, the dual derivative is not a good object because
\begin{equation}
 \delta_\xi \tpartial^{\alpha\beta} \phi \neq L_\xi \tpartial^{\alpha\beta} \phi \,,
\end{equation}
for a scalar $\phi$. However, we can define an improved dual derivative
\begin{equation}
 \hpartial^{\alpha\beta} \phi = \tpartial^{\alpha\beta} \phi + \Omega^{\alpha\beta\gamma} \partial_\gamma \phi \,, \label{eq:ImprovedWrapping}
\end{equation}
which does transform covariantly.

\paragraph{M-theory R-flux}
We now wish to generalise \eqref{eq:StringRflux} using the dual derivatives $\tpartial^{\alpha\beta}$ (or their improved version $\hpartial^{\alpha\beta}$) and the trivector $\Omega^{\alpha\beta\gamma}$. It is clear that the $R$-flux cannot be
\begin{equation}
 R^{\alpha\beta\gamma\delta\rho} \neq 5 \hpartial^{[\alpha\beta} \Omega^{\gamma\delta\rho]} \,,
\end{equation}
since $\alpha, \beta = 1, \ldots, 4$ and so this vanishes identically. However, the following object is a spacetime tensor.
\begin{equation}
 R^{\alpha,\beta\gamma\delta\rho} = 4 \hpartial^{\alpha[\beta} \Omega^{\gamma\delta\rho]} \,,
\end{equation}
and is the generalisation of the $R$-flux for M-theory. To show this, first note that the section condition is now \cite{Berman:2011cg}
\begin{equation}
 \partial_{[ab} f \partial_{cd]} g = \partial_{[ab} \partial_{cd]} g = 0 \,,
\end{equation}
where $a, b = 1, \ldots, 5$ and $\partial_{ab} = \partial_{[ab]}$ are the 10 generalised derivatives. In particular, the spacetime derivative is
\begin{equation}
 \partial_\alpha = \partial_{\alpha 5} \,,
\end{equation}
and the dual derivative is
\begin{equation}
 \tpartial^{\alpha\beta} = \frac12 \epsilon^{\alpha\beta\gamma\delta} \partial_{\gamma\delta} \,,
\end{equation}
where $\epsilon^{\alpha\beta\gamma\delta} = \pm 1$ is the alternating tensor density. Thus, the section condition means in particular that
\begin{equation}
 \tpartial^{[\alpha\beta} f \tpartial^{\gamma\delta]} g = \tpartial^{[\alpha\beta} \tpartial^{\gamma\delta]} f = 0 \,.
\end{equation}
Now consider
\begin{equation}
 \begin{split}
  \delta_\xi \hpartial^{\alpha\beta} \Omega^{\gamma\delta\rho} &= L_\xi \hpartial^{\alpha\beta} \Omega^{\gamma\delta\rho} - 3 \Omega^{p[\delta\rho} \hpartial^{|\alpha\beta|} \partial_p \xi^{\gamma]} - 3 \hpartial^{\alpha\beta} \partial^{[\gamma\delta} \xi^{\rho]} \\
  &= L_\xi \hpartial^{\alpha\beta} \Omega^{\gamma\delta\rho} - 3 \Omega^{\sigma[\delta\rho} \tpartial^{|\alpha\beta|} \partial_\sigma \xi^{\gamma]} - \Omega^{\tau[\delta\rho} \Omega^{|\alpha\beta\sigma|} \partial_\sigma \partial_\tau \xi^{\gamma]} - 3 \tpartial^{\alpha\beta} \partial^{[\gamma\delta} \xi^{\rho]} \\
  & \quad - 3 \Omega^{\alpha\beta\sigma} \partial_\sigma \tpartial^{[\gamma\delta} \xi^{\rho]} \\
  &= L_\xi \hpartial^{\alpha\beta} \Omega^{\gamma\delta\rho} - 3 \Omega^{\sigma[\delta\rho} \tpartial^{|\alpha\beta|} \partial_\sigma \xi^{\gamma]} - 3 \Omega^{\sigma\alpha\beta} \tpartial^{[\gamma\delta} \partial_\sigma \xi^{\rho]} - 3 \tpartial^{\alpha\beta} \tpartial^{[\gamma\delta} \xi^{\rho]} \\
  & \quad - 3 \Omega^{\sigma\alpha\beta} \Omega^{\tau[\gamma\delta} \partial_\sigma \partial_\tau \xi^{\sigma]} \,.
 \end{split}
\end{equation}
When we antisymmetrise over $\beta\gamma\delta\rho$, the second and third anomalous terms become
\begin{equation}
 -3 \Omega^{\sigma[\delta\rho} \tpartial^{|\alpha|\beta} \partial_\sigma \xi^{\gamma]} - 3 \Omega^{\sigma\alpha[\beta} \tpartial^{\gamma\delta} \partial_\sigma \xi^{\rho]} \propto \Omega^{\sigma[\delta\rho} \tpartial^{\alpha\beta} \partial_\sigma \xi^{\gamma]} = 0 \,,
\end{equation}
where we have used the fact that $\alpha, \beta = 1, \ldots, 4$ and so an antisymmetrisation over 5 indices vanishes. Similarly, the fourth anomalous term becomes
\begin{equation}
 - 3 \tpartial^{\alpha[\beta} \tpartial^{\gamma\delta} \xi^{\rho]} \propto \tpartial^{[\alpha\beta} \tpartial^{\gamma\delta} \xi^{\rho]} = 0 \,.
\end{equation}
This relies on the fact that $\tpartial^{\alpha\beta} \tpartial^{\gamma\delta}$ is symmetric on the interchange of derivatives. Similarly the final term vanishes upon the appropriate antisymmetrisation since
\begin{equation}
 \Omega^{\sigma\alpha[\beta} \Omega^{|\tau|\gamma\delta} \partial_\sigma \partial_\tau \xi^{\rho]} \propto \Omega^{\sigma[\alpha\beta} \Omega^{|\tau|\gamma\delta} \partial_\sigma \partial_\tau \xi^{\rho]} = 0 \,.
\end{equation}
Hence
\begin{equation}
 \delta_\xi \hpartial^{\alpha[\beta} \Omega^{\gamma\delta\rho]} = L_\xi \hpartial^{\alpha[\beta} \Omega^{\gamma\delta\rho]} \,,
\end{equation}
and $R^{\alpha,\beta\gamma\delta\rho} = 4 \hpartial^{\alpha[\beta} \Omega^{\gamma\delta\rho]}$ is a spacetime tensor.

\subsection{Generalised vielbein of $Q$-flux background} \label{s:DFTQFluxBackgrounds}
As noted in \cite{Grana:2008yw,Andriot:2012an} the geometric frame is not globally well-defined for a non-geometric background. In the toy model given this can be seen immediately by studying the generalised vielbein.

The vielbein can be written in terms of the geometric frame variables as
\begin{equation}
 E_I{}^{\bar{I}} = \begin{pmatrix}
  e_i{}^{\bar{i}} & B_{ij} e^{j}{}_{\bar{i}} \\
  0 & e^i{}_{\bar{i}}
 \end{pmatrix} \,,
\end{equation}
or in terms of the non-geometric frame variables as
\begin{equation}
 E_I{}^{\bar{I}} = \begin{pmatrix}
  e_i{}^{\bar{i}} & 0 \\
  \beta^{ij} e_j{}^{\bar{i}} & e^i{}_{\bar{i}}
 \end{pmatrix} \,,
\end{equation}
where $I = 1, \ldots, 2D$ are the $\mathrm{O}(D,D)$ indices, $e_i{}^{\bar{i}}$ is the vielbein of the spacetime metric, $B_{ij}$ is the Kalb-Ramond form and $\beta^{ij}$ is the aforementioned bivector.\footnote{One can also consider more general parameterisations of the generalised vielbein including both $B$ and $\beta$ but we will not need to consider this for our purposes.}

If we now revisit the twisted torus, we can use the globally well-defined 1-forms $\eta^1, \eta^2, \eta^3$ of \eqref{eq:TwistedTorus1Forms} to write the generalised vielbein of the twisted torus as
\begin{equation}
 \left(E^{TT}\right)_A{}^{\bar{A}} = \begin{pmatrix}
  1 & 0 & 0 & 0 & 0 & 0 \\
  - N \z & 1 & 0 & 0 & 0 & 0 \\
  0 & 0 & 1 & 0 & 0 & 0 \\
  0 & 0 & 0 & 1 & 0 & 0 \\
  0 & 0 & 0 & 0 & 1 & 0 \\
  0 & 0 & 0 & 0 & 0 & 1
 \end{pmatrix} \,,
\end{equation}
where the superscript ``$TT$'' stands for twisted torus. This generalised vielbein is globally well-defined since the space is parallelisable (and generalised parallelisable). After the duality $T^2$ along the direction $\y$ we obtain the generalised vielbein for the $Q$-flux background
\begin{equation}
 \left(E^R\right)_A{}^{\bar{A}} = \begin{pmatrix}
  1 & 0 & 0 & 0 & 0 & 0 \\
  0 & 1 & 0 & 0 & 0 & 0 \\
  0 & 0 & 1 & 0 & 0 & 0 \\
  0 & 0 & 0 & 1 & 0 & 0 \\
  - N \z & 0 & 0 & 0 & 1 & 0 \\
  0 & 0 & 0 & 0 & 0 & 1
 \end{pmatrix} \,.
\end{equation}
We can now immediately see that the parameterisation with the bivector is globally well-defined and from \eqref{eq:QFluxDef} we find $Q_3^{12} = N$.

\section{Octonions, Malcev Algebras and their deformations}
\label{s:AppOctonions}

\subsection{Octonions, their multiplication table and quaternion subalgebras} \label{s:AppOctMult}
\begin{figure}
\centering
  \includegraphics{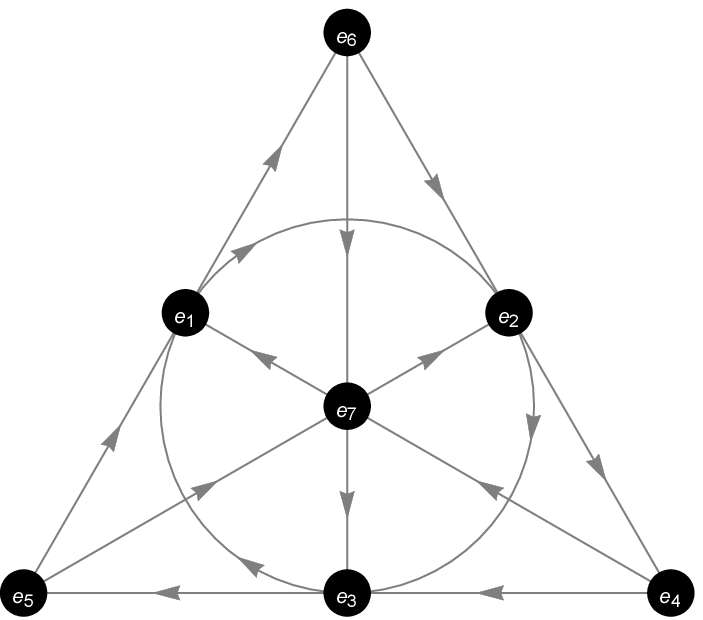}
  \caption{Multiplication table of imaginary units of real octonions $\mathbb{O}$. The three imaginary units on each side, height and circle correspond to the imaginary units of a quaternion subalgebra. The arrows represent the positive directions for multiplication, e.g. $e_1 e_2 =-e_2 e_1=e_3 $ and $e_6e_2=-e_2e_6=e_4$, etc..}\label{f:fano}
\end{figure}

A composition algebra $\mathbb{A}$ is a finite-dimensional algebra with identity that is endowed with a quadratic norm $Q$ that satisfies the property 
\begin{equation}
Q(XY) =Q(X)\, Q(Y) \quad \forall \quad X,Y\in \mathbb{A} \,.
\end{equation} 
If every non-zero element has an inverse in the algebra $\mathbb{A}$ it is called a division algebra. There exist four composition algebras over the field of real numbers, namely the real numbers $\mathbb{R}$, complex numbers $\mathbb{C}$ , quaternions $\mathbb{H}$ and octonions $\mathbb{O}$. The division algebra of real octonions $\Oct$ is a non-commutative and non-associative algebra with seven imaginary units $e_A \, (A,B =1,\ldots, 7)$ that satisfy
\begin{equation}
 e_A e_B = -\delta_{AB} + \eta_{ABC} \, e_C \,,
\end{equation}
where $\eta_{ABC}$ are the completely anti-symmetric structure constants. The non-vanishing components of $\eta_{ABC}$, in the conventions of \cite{Gunaydin:1973rs}, are given by
\begin{equation}
 \eta_{ABC} =1 \Longleftrightarrow   (ABC) = (123),\, (516),\, (624),\, (435),\, (471),\, (572),\, (673)\,,
\end{equation}
and cyclic permutations thereof. The multiplication table can be conveniently represented as in figure \ref{f:fano}, where the three imaginary units along each side, each perpendicular and the units $e_1,e_2 $ and $e_3$ along the circle belong to a quaternion subalgebra which is non-commutative, but associative.

The associator of any three elements $X,Y$ and $Z$ of $\Oct$ is defined as
\begin{equation}
 [X,Y,Z] \equiv (XY) Z - X(YZ) \,,
\end{equation}
which satisfies
\begin{equation}
[X,Y,Z]=[Z,X,Y]=[Y,Z,X]=-[Y,X,Z] \,.
\end{equation}
An octonion can be represented as a pair of quaternions as follows
\begin{equation}
X = X_0 + X_A e_A = ( X_0 + X_i e_i ) + e_7 ( X_7 + X_{(i+3)} e_i ) \,,
\end{equation}
where $i,j=1,2,3$ and we used the fact that $e_{(i+3)} = e_7 e_i$. Its automorphism group is the exceptional group $G_2$ and the invariance group of the norm defined as
\begin{equation}
 Q(X) \equiv X \bar{X} = X_0^2 + X_A X_A \,,
\end{equation}
is $\mathrm{SO}{(8)}$, where the conjugate octonion $\bar{X}$ is obtained by replacing all the imaginary units $e_A$ by their negatives
\begin{equation}
 \bar{X} = X_0 - X_A e_A \,.
\end{equation}
A real octonion $X$ can be written as a pair of quaternions as follows:
\begin{equation}
X= X_0 + X_i e_i + e_7 ( X_7 + X_{i+3} e_i ) \,,
\end{equation}
where $ e_i \, ( i=1,2,3)$ are the imaginary units of a quaternion subalgebra. Split octonions $\mathbb{O}^S$ on the other hand do not form a division algebra. A split octonion $X^s$ can be expanded as follows \cite{Gunaydin:1973rs} 
\begin{equation}
 X^s = X_0 + X_i e_i + i e_7 ( X_7 + X_{i+3} e_i ) =X_0+X_i e_i + i e_7 + i e_{i+3} X_{i+3} \,,
\end{equation}
where $i$ is an imaginary unit that commutes with $e_A$. The norm of  split octonion $X^s$ is given by
\begin{equation}
Q(X^s) = X^s \bar{X}^s = X_0^2 + X_1^2 + X_2^2 +X_3^2  - (X_4^2+X_5^2+X_6^2+X_7^2) \,,
\end{equation}
where $\bar{X}^s=  X_0 - X_i e_i - i e_7 ( X_7 + X_{i+3} e_i ) $ and 
whose invariance group is $\mathrm{SO}(4,4)$. The automorphism group of split octonions $\mathbb{O}^S$ is the noncompact split $G_{2(2)}$ with the maximal compact subgroup $\SOf$.

\subsection{Octonions and Malcev Algebras} \label{s:AppMalcev}
A Malcev algebra is an algebra with an anti-symmetric product
\[ a \star b  = - b \star a \,, \]
that satisfies the Malcev identity
\[ (a \star b) \star (a \star c) = ((a \star b) \star c) \star a + ((b \star c) \star a) \star a + ((c \star a) \star a ) \star b \,. \]
The Malcev identity can be rewritten  in the form
\[ J(a,b,a \star c) = J(a,b,c) \star a \,, \]
where $J(a,b,c)$ is the Jacobiator
\[ J(a,b,c) \equiv ((a\star b)\star c) + (( c \star a) \star b) + (( b \star c) \star a) \,. \]

The imaginary units $e_A$ of octonions form a simple Malcev algebra under the commutator product
\begin{equation}
 e_A \star e_B \equiv [ e_A, e_B] \,.
\end{equation}

\section{$\SOf$-invariance of the locally non-geometric M-theory algebra}
\label{s:AppSO4}
Upon first seeing the algebra \eqref{eq:FullMAlgebra} one may wonder why it is not invariant under $\textrm{GL}(4)$, just as the string $R$-flux algebra is invariant under $\textrm{GL}(3)$. This would normally be interpreted as ``coordinate invariance''. However, for the M-theory $R$-flux background there is a preferred coordinate choice where $X^4$ is singled out as the coordinate with no dual momentum. This can be seen from the M-theory $R$-flux tensor which transforms as a vector under $\textrm{GL}(4)$ and thus breaks the $\textrm{GL}(4)$ symmetry.

However, there is a three-dimensional representation of $\SOf$ and under $\SOf \simeq \SUt \times \SUt / \mathbb{Z}_2$ the coordinates transform in the $\left(\mathbf{2},\mathbf{2}\right)$ representation of $\SUt \times \SUt$ while the momenta transform in the $\left(\mathbf{3}, \mathbf{1}\right)$ representation. We can thus write the algebra in a manifestly $\SOf$-invariant way, as we will now do.\footnote{Alternatively, and equivalently, one could make the $\SOf$-invariance manifest by identifying the three momenta with self-dual two-forms of $\SOf$ but here we prefer to work with $\SUt\times\SUt$.}
It is important to note that the $\SOf$ symmetry does not act consistently on $R^{\alpha,\beta\gamma\delta\rho}$ since it leaves $R^{4,\alpha\beta\gamma\delta} = \epsilon^{\alpha\beta\gamma\delta}$ invariant.

\subsection{$\SOf \sim \SUt \times \SUt / \mathbb{Z}_2$ conventions}
\label{s:AppSO4Conventions}
We begin by introducing the $\SOf$ gamma matrices
\begin{equation}
 \Gamma^{(\alpha} \bar{\Gamma}^{\beta)} = \delta^{\alpha\beta} \mathbf{1} \,, \qquad \bar{\Gamma}^{(\alpha} \Gamma^{\beta)} = \delta^{\alpha\beta} \mathbf{1} \,,
\end{equation}
where
\begin{equation}
 \left(\Gamma^\alpha\right)_{a\dot{b}} = \left( -i\sigma^i, \mathbf{1} \right)_{a\dot{b}} \,,
\end{equation}
and
\begin{equation}
 \left( \bar{\Gamma}^{\alpha} \right)^{\dot{a}b} = \epsilon^{\dot{a}\dot{c}} \epsilon^{bd} \left(\Gamma^\alpha\right)_{\dot{c}d} = \left( i \sigma^i, \mathbf{1} \right)^{\dot{a}b} \,.
\end{equation}
Here $a = 1, 2$ are $\SUt_L$ indices, $\dot{a} = \dot{1}, \dot{2}$ are $\SUt_R$ indices, $\sigma^i$ are the Pauli matrices
\begin{equation}
 \sigma^1 = \begin{pmatrix}
  0 & 1 \\ 1 & 0
 \end{pmatrix} \,, \qquad
 \sigma^2 = \begin{pmatrix}
  0 & -i \\ i & 0
 \end{pmatrix}\,, \qquad
 \sigma^3 = \begin{pmatrix}
  1 & 0 \\ 0 & -1
 \end{pmatrix}\,,
\end{equation}
and $\mathbf{1}$ denotes the $2\times 2$ unit matrix.

We use these gamma matrices to map the $\mathbf{4}$ of $SO(4)$ to the $\left(\mathbf{2},\mathbf{2}\right)$ of $SU(2)\times SU(2)$.
\begin{equation}
 X^{\dot{a}b} = \left(\bar{\Gamma}_\alpha\right)^{\dot{a}b} X^\alpha \,, \qquad X_{a\dot{b}} = \left(\Gamma_\alpha\right)_{a\dot{b}} X^\alpha \,, \qquad X^\alpha = \frac12 \left(\Gamma^\alpha\right)_{a\dot{b}} X^{\dot{b}a} = \frac12 \left(\bar{\Gamma}^\alpha\right)^{\dot{a}b} X_{b\dot{a}} \,. \label{eq:Xadotb}
\end{equation}
Similarly we can write the momenta, which transform in the $\left(\mathbf{3},\mathbf{1}\right)$, as
\begin{equation}
 P_a{}^b = \left(\Gamma_i\right)_{a\dot{c}} \left( \bar{\Gamma}^4\right)^{\dot{c}b} P^i \,, \qquad P^i = \frac12 \left(\Gamma^4\right)_{b\dot{c}} \left( \bar{\Gamma}^i \right)^{\dot{c}a} P_a{}^b \,, \label{eq:Pab}
\end{equation}
with $P_a{}^a = 0$.

Raising and lowering indices is done with the $\SUt$-invariant tensors $\epsilon_{ab}$ and $\epsilon_{\dot{a}\dot{b}}$, where we use the convention that
\begin{equation}
 P_{ab} = P_a{}^c \epsilon_{cb} \,, \qquad P^{ab} = \epsilon^{ac} P_c{}^b \,.
\end{equation}

For calculations the completeness relation for the gamma matrices
\begin{equation}
 \left(\bar{\Gamma}_\alpha\right)^{\dot{a}b} \left(\Gamma^\alpha\right)_{c\dot{d}} = 2 \delta^{\dot{a}}_{\dot{d}} \delta^b_c \,,
\end{equation}
and
\begin{equation}
 2 \epsilon_{\alpha\beta\gamma\delta} = \mathrm{tr}\left( \Gamma_{\alpha}\bar{\Gamma}_\beta \Gamma_{[\gamma} \bar{\Gamma}_{\delta]} \right) + 4 \delta_{\alpha[\gamma} \delta_{\delta]\beta} \,,
\end{equation}
are useful.

\subsection{The $\SOf$-invariant non-associative algebra}
\label{s:AppSO4Manifest}
Using the conventions outlined above we can write the commutators of \eqref{eq:FullMAlgebra} as
\begin{equation}
 \begin{split}
  \left[ P_a{}^b, P_c{}^d \right] &= - 2 i \hbar \left(P_{[a}{}^{(b} \delta^{d)}_{c]} - P_{(a}{}^{[b} \delta^{d]}_{c)} \right) \,, \\
  \left[ P_a{}^b, X_{c\dot{c}} \right] &= 2 i \hbar \left( \delta_c^b X_{a\dot{c}} - \frac12 \delta_a^b X_{c\dot{c}} \right) \,, \\
  \left[ X_{a\dot{a}}, X_{b\dot{b}} \right] &= - \frac{2iNl_s^3}{\hbar} \epsilon_{\dot{a}\dot{b}} P_{ab} \,, \label{eq:MCommR}
 \end{split}
\end{equation}
and the associators as
\begin{equation}
 \begin{split}
  \left[ P_a{}^b, X_{c\dot{c}}, X_{d\dot{d}} \right] &= - 4 N l_s^3  \epsilon_{\dot{c}\dot{d}} \left( \epsilon_{a(c} P_{d)}{}^b - \frac12 \delta_a^b P_{dc} \right) \,, \\
  \left[ P_a{}^b, P_c{}^d, X_{e\dot{e}} \right] & = - 4 \hbar^2 \left(\epsilon_{ac} \delta_{e}^{(b} X^{d)}{}_{\dot{e}} + \epsilon^{bd} \epsilon_{e(a} X_{c)\dot{e}} \right) \,, \\
  \left[ X_{a\dot{a}}, X_{b\dot{b}}, X_{c\dot{c}} \right] &= 4 N l_s^3 \left( X_{b[\dot{c}} \epsilon_{\dot{b}]\dot{a}} \epsilon_{ac} + X_{a[\dot{a}} \epsilon_{\dot{c}]\dot{b}} \epsilon_{bc} \right) \,, \\
  \left[ P_a{}^b, P_c{}^d, P_e{}^f \right] &= 0 \,. \label{eq:MAssocR}
 \end{split}
\end{equation}
Up to the overall coefficients, the right-hand-sides of these equations are uniquely fixed by requiring $\SUt \times \SUt$ invariance.

\bibliographystyle{JHEP}
\bibliography{NewBib}
\end{document}